\documentclass[default,iicol,pdflatex,sn-nature]{sn-jnl}

\usepackage{graphicx}%
\usepackage{multirow}%
\usepackage{amsmath,amssymb,amsfonts}%
\usepackage{amsthm}%
\usepackage{mathrsfs}%
\usepackage[title]{appendix}%
\usepackage{xcolor}%
\usepackage{textcomp}%
\usepackage{manyfoot}%
\usepackage{booktabs}%
\usepackage{algorithm}%
\usepackage{algorithmicx}%
\usepackage{algpseudocode}%
\usepackage{listings}%
\usepackage{geometry}
\usepackage{fancyhdr}
\usepackage{lineno}

\begin{document}

\title[Beam Shaping by Nonlinear Moir\'e Metasurfaces]{Beam shaping by nonlinear moir\'e metasurfaces}

\author[1]{Lun Qu}
\equalcont{These authors contributed equally to this work.}

\author[1]{Wei Wu}
\equalcont{These authors contributed equally to this work.}

\author[1,2]{Di Zhang}
\author[1]{Chenxiong Wang}
\author[1]{Lu Bai}
\author[1]{Chenyang Li}
\author[1]{Wei Cai}
\author*[1,4]{Mengxin Ren} \email{ren\_mengxin@nankai.edu.cn}
\author*[2,3]{Andrea Al\`u} \email{aalu@gc.cuny.edu}
\author*[1]{Jingjun Xu} \email{jjxu@nankai.edu.cn}

\affil[1]{\orgdiv{The Key Laboratory of Weak-Light Nonlinear Photonics, Ministry of Education, School of Physics and TEDA Applied Physics Institute}, \orgname{Nankai University}, \orgaddress{\city{Tianjin}, \postcode{300071}, \country{China}}}

\affil[2]{Photonics Initiative, Advanced Science Research Center, City University of New York, New York, NY 10031, USA}

\affil[3]{Physics Program, Graduate Center, City University of New York, New York, NY 10016, USA}

\affil[4]{\orgdiv{Collaborative Innovation Center of Extreme Optics}, \orgname{Shanxi University}, \orgaddress{\city{Taiyuan, Shanxi}, \postcode{030006}, \country{China}}}

\abstract{This paper explores the interplay of momentum transfer and nonlinear optical processes through moir\'e phenomena. Momentum transfer plays a crucial role in the interaction between photons and matter. Here, we study stacked metasurfaces with tailored dispersion and rotated against each other with varying twisted angles. The stacking introduces interlayer interactions, which can be controlled by the relative angle between metasurfaces, significantly enriching the resulting response compared to the single layer counterpart. By focusing on second-harmonic generation (SHG) from these twisted metasurfaces, we delve into the realm of nonlinear moir\'e photonics. Through experimental observations, we unveil the emergence of intricate far-field SHG radiation patterns, showing their effective tuning by varying the twisted angles. These findings offer a fresh perspective to explore nonlinear wavefront shaping through moir\'e phenomena, opening new avenues for nonlinear information processing, optical steering, and nonlinear optical switching.}


\maketitle

\section{Introduction}
Momentum transfer plays an important role in governing the interaction between photons and matter.\textsuperscript{\cite{mobley2018momentum}} This effect can be observed in the diffraction from periodic structures such as gratings, where momenta associated with the reciprocal lattice vectors are transferred to photons, altering the direction of light propagation.\textsuperscript{\cite{xu2023line}} The concept of momentum transfer is essential in the design of optical elements that control and manipulate light, facilitating applications such as beam steering,\textsuperscript{\cite{liu2017beam, wei2018experimental,tang2019quasicrystal}} beam splitting,\textsuperscript{\cite{ra2017metagratings}} information processing,\textsuperscript{\cite{ren2020complex}} and multichannel communications.\textsuperscript{\cite{zhao2018multichannel}} It forms the foundation for advancing optical technologies with wide-ranging implications across diverse scientific and engineering domains.

The quest for enhanced flexibility in light manipulation necessitates the establishment of highly reconfigurable devices capable of yielding a customizable spectrum of momentum components. Moir\'e phenomena, arising from the superposition of two or more periodic lattices with twisted angles or mismatched lattice constants, emerge as a versatile platform to generate diverse lattices with variable symmetries and periodicities.\textsuperscript{\cite{lau2022reproducibility, du2023moire, sunku2018photonic, lu2018valley, hu2020topological, yang2020tunable, yao2021enhanced, kim2023three, hu2021tailoring, chen2021perspective, li2024twisted, zhou2024dynamic}} This large degree of tunability is achieved through the manipulation of the twist angle and of the relative lattice constant, providing a promising avenue for the creation of new Brillouin zones with intricate band structures and a rich manipulation of momentum space.\textsuperscript{\cite{amidror2003moire, lubin2012high, baake2012mathematical, lou2021theory, lou2022tunable1, tang2023experimental}} The control of the twisted angles facilitates band structure manipulation, leading to phenomena such as remarkably flat photonic bands,\textsuperscript{\cite{lopez2020flat, deng2020magic, dong2021flat, tang2021modeling, oudich2021photonic, yi2022strong, huang2022moire}} light localization-delocalization transitions,\textsuperscript{\cite{wang2020localization, fu2020optical, nguyen2022magic}} and photonic band structure engineering.\textsuperscript{\cite{wang2018bound, hu2020moire, lou2021theory, lou2022tunable1, lou2022tunable2, salakhova2023twist, tang2023experimental}}

In a parallel effort, metasurfaces, renowned for their control over light wavefronts, serve as a captivating framework for the design of both real and reciprocal lattices. The fusion of metasurface concepts with moir\'e photonics has resulted in substantial advancements in nanophotonics. This field has witnessed notable progress and has catalyzed innovative applications, including the development of photonic magic-angle nanolasers,\textsuperscript{\cite{mao2021magic, raun2023gan, luan2023reconfigurable}} and steering of both microwave and optical laser emissions.\textsuperscript{\cite{liu2022moire, guan2023far, guan2023plasmonic}} Additionally, optical vortex generation and enhanced optical chirality have been demonstrated from twisted moir\'e metasurfaces.\textsuperscript{\cite{zhang2023twisted, wu2017moire, han2023recent, zhang2023nanoimprint}} Despite these significant advances, nonlinear phenomena in moir\'e metasurfaces have not been experimentally explored yet. Yet, the new degree of control over dispersion and momentum offered by moir\'e metasurfaces can provide exciting opportunities for nonlinear optical modulation.

In this paper, we delve into the capabilities of moir\'e metasurfaces to govern nonlinear optical processes, with a specific focus on the manipulation of second harmonic generation (SHG). Our experimental approach involves the fabrication of stacked nonlinear moir\'e metasurfaces with varied twisted angles. The interlayer interaction in these structures results in a combination of reciprocal lattice vectors from the two elementary metasurfaces, thereby enriching the momentum components compared to their single layer counterparts. The intricate far-field SHG radiation patterns observed in our experiments are efficiently tuned by adjusting the twisted angles. The intriguing possibility of shaping SHG with moir\'e phenomena adds new versatility to nonlinear photonics, opening avenues for innovative applications in information processing, optical routing, as well as nonlinear optical switching and modulation. Our findings also provide valuable opportunities for the modulation of nonlinear wavefronts and diffraction, paving the way to further exploration of nonlinear phenomena within the realm of moir\'e and aperiodic photonics.

\section{Results}
\textbf{Figure 1} shows a conceptual diagram of a stacked nonlinear moir\'e metasurface designed to manipulate SHG radiation. The chosen nonlinear constitutive material is $x$-cut lithium niobate (LN), renowned for its large second-order nonlinearity and broad transparent window.\textsuperscript{\cite{weis1985lithium, ma2020second, jia2021ion}} This material is commonly employed in the study of nonlinear metasurfaces for frequency conversion.\textsuperscript{\cite{ma2021nonlinear, qu2022giant, fang2020second, fedotova2020second, carletti2021Steering, yuan2021strongly, huang2022resonant, zhang2022spatially, qu2023bright, zhao2023efficient, fedotova2022lithium}} In our design, the two metasurface layers are composed of a square lattice of circular air holes, stacked with a relative twisted angle $\alpha$, forming a distinct moir\'e pattern in the overlapped region. When excited by a fundamental frequency (FF) beam, nonlinear polarization dipoles $P^{(2)}$ are generated inside the LN via the second-order susceptibility $\chi^{(2)}$. These dipoles oscillate at twice the frequency of the FF fields, acting as secondary sources that radiate second harmonic (SH) waves to the far-field. Each metasurface hole array forms a nonlinear photonic lattice with a periodic distribution of $\chi^{(2)}$, which gives rise to reciprocal lattice vectors and corresponding lattice momentum perpendicular to the FF beam. In addition to the intrinsic lattice momentum of each metasurface, the quasi-periodic moir\'e lattice supports additional momentum components resulting from the interlayer coupling between the stacked layers. These momenta transfer to SHG photons and modulate the propagation direction of the SHG waves.

\begin{figure}[H]
  \centering
  \includegraphics[width=95mm]{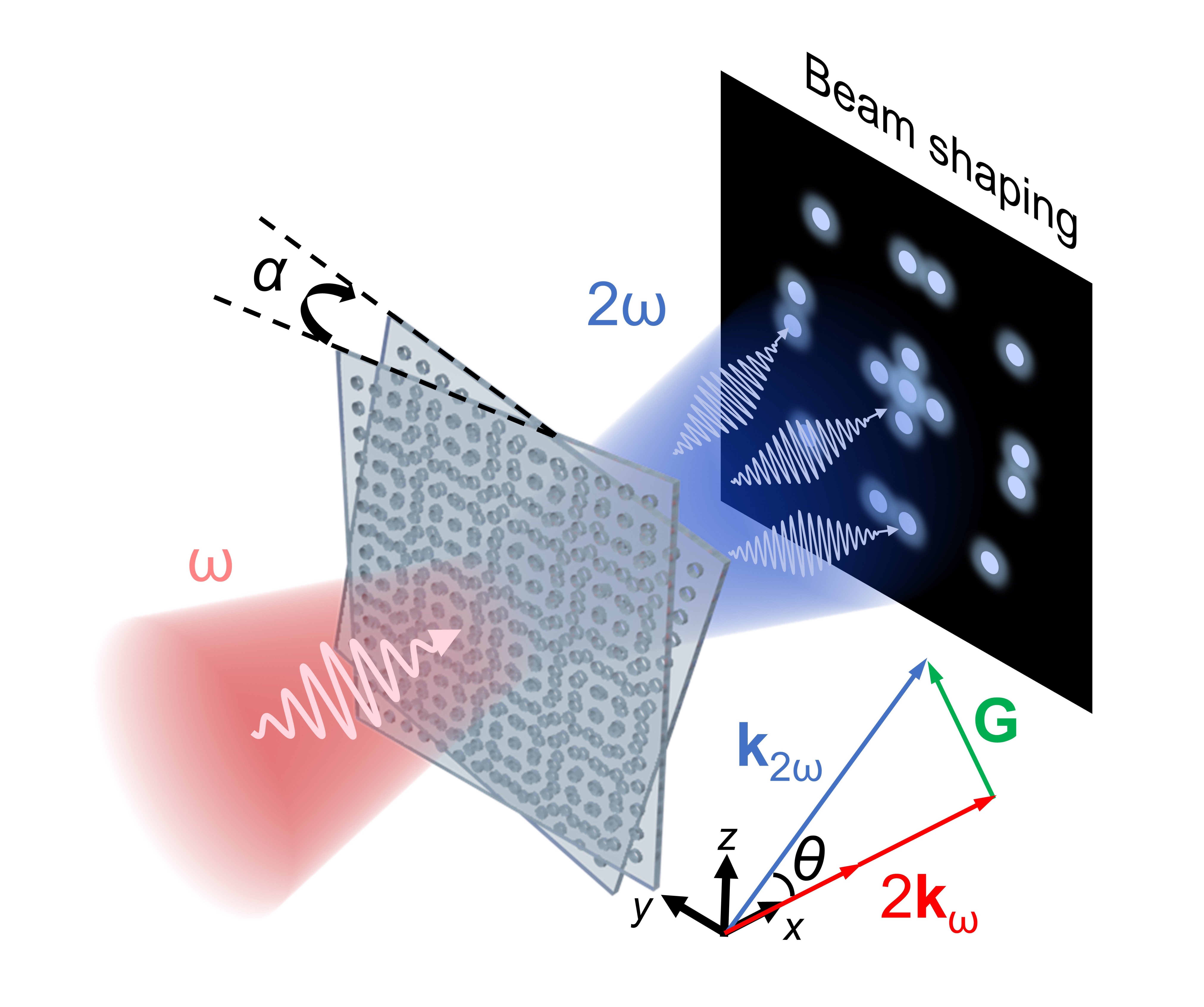}
  \caption{Schematic representation of stacked nonlinear moir\'e metasurfaces to control SHG radiation. The moir\'e metasurface comprises two layers, each featuring a square lattice of hole arrays milled in an $x$-cut LN membrane. The experimental coordinate was chosen to overlap with the principal crystallographic axes of the untwisted layer LN, defining $z$-axis along its optic axis ($e$). The other layer introduces a counterclockwise twist by an angle $\alpha$ (as viewed against the direction of incident light). The FF light beam is normally incident onto the structure, while SHG signals are collected in the transmission direction. A sketch of momentum transfer is shown in the bottom-right corner. The red and blue arrows represent the wave vectors of the FF light and SHG wave, respectively. The green arrow denotes the reciprocal lattice vector of the metasurface. The emission angle $\theta$ of the SHG is defined as the angle between the SHG beam path and the normal of the metasurface.}
  \label{fig1}
\end{figure}
\subsection{Fabrication of bilayer moir\'e metasurfaces}

\begin{figure*}[htbp]
  \centering
  \includegraphics[width=160mm]{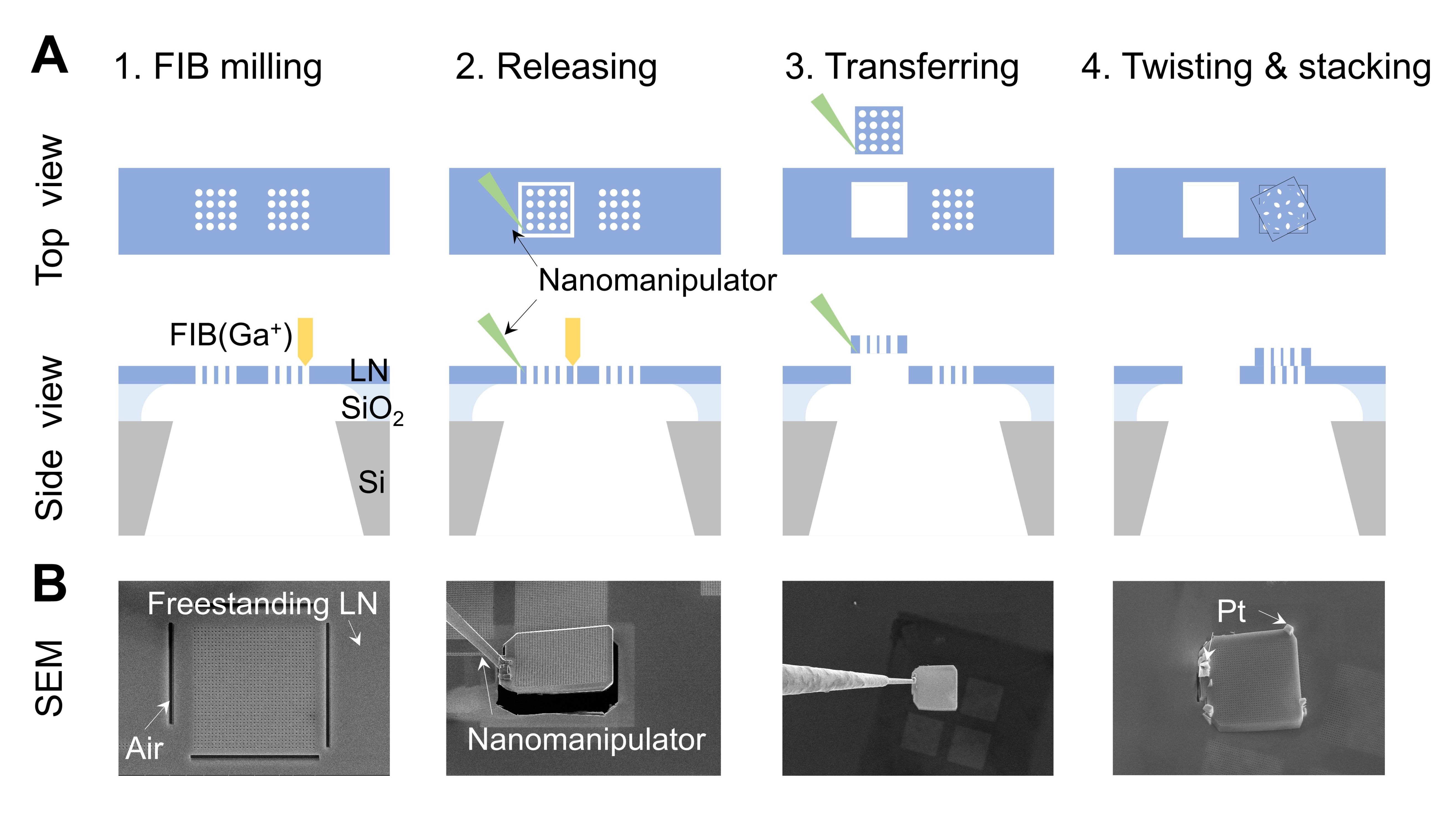}
  \caption{Fabrication process of stacked moir\'e metasurfaces. a) Depiction of key steps in fabricating bilayer moir\'e metasurfaces, presented in both top and side views. The process involves precise techniques of FIB milling and nanomanipulator stacking operation. b) Corresponding SEM images providing visual representations of the described fabrication steps, which confirm the successful realization of the designed moir\'e metasurfaces.}
  \label{fig2} 
\end{figure*}

Our nonlinear moir\'e metasurfaces were fabricated in a 210~nm thick $x$-cut single-crystalline freestanding LN membrane.\textsuperscript{\cite{qu2022giant}} The detailed procedures for fabricating the freestanding LN membrane are outlined in the Experimental Section. \textbf{Figure 2} illustrates the step-by-step fabrication process for the stacked moir\'e metasurfaces. Initially, we employed a focused ion beam (FIB, Ga$^+$ ion, 30 kV, 24 pA) milling technique for metasurface fabrication. The air holes had a diameter ($D$) of 225~nm and the lattice constant ($P$) was 600~nm (Figure S2a, Supporting Information). Seven metasurface arrays were fabricated, each occupying a 25 $\times$ 25~$\mu$m$^2$ area. To stack one LN metasurface onto another, we first partially cut one metasurface using the ion beam. Subsequently, we welded a nanomanipulator needle tip to the metasurface using ion-assisted platinum (Pt) deposition. The remaining connecting sections were milled by ions to completely release the metasurface from the membrane. The lower membrane was twisted by a specific angle by rotating the sample holder. During stacking, the Pt deposition was used to bond the overlayer metasurface to the bottom metasurface at corners. Finally, the needle tip was cut away by ion milling, fully releasing the metasurface and completing the assembly of the bilayer moir\'e metasurfaces (Figure S2b, Supporting Information). Figure 2b provides a visual depiction of the fabrication process through scanning electron microscope (SEM) images.

\subsection{Optical moir\'e patterns}

\begin{figure*}[htbp]
  \centering
  \includegraphics[width=160mm]{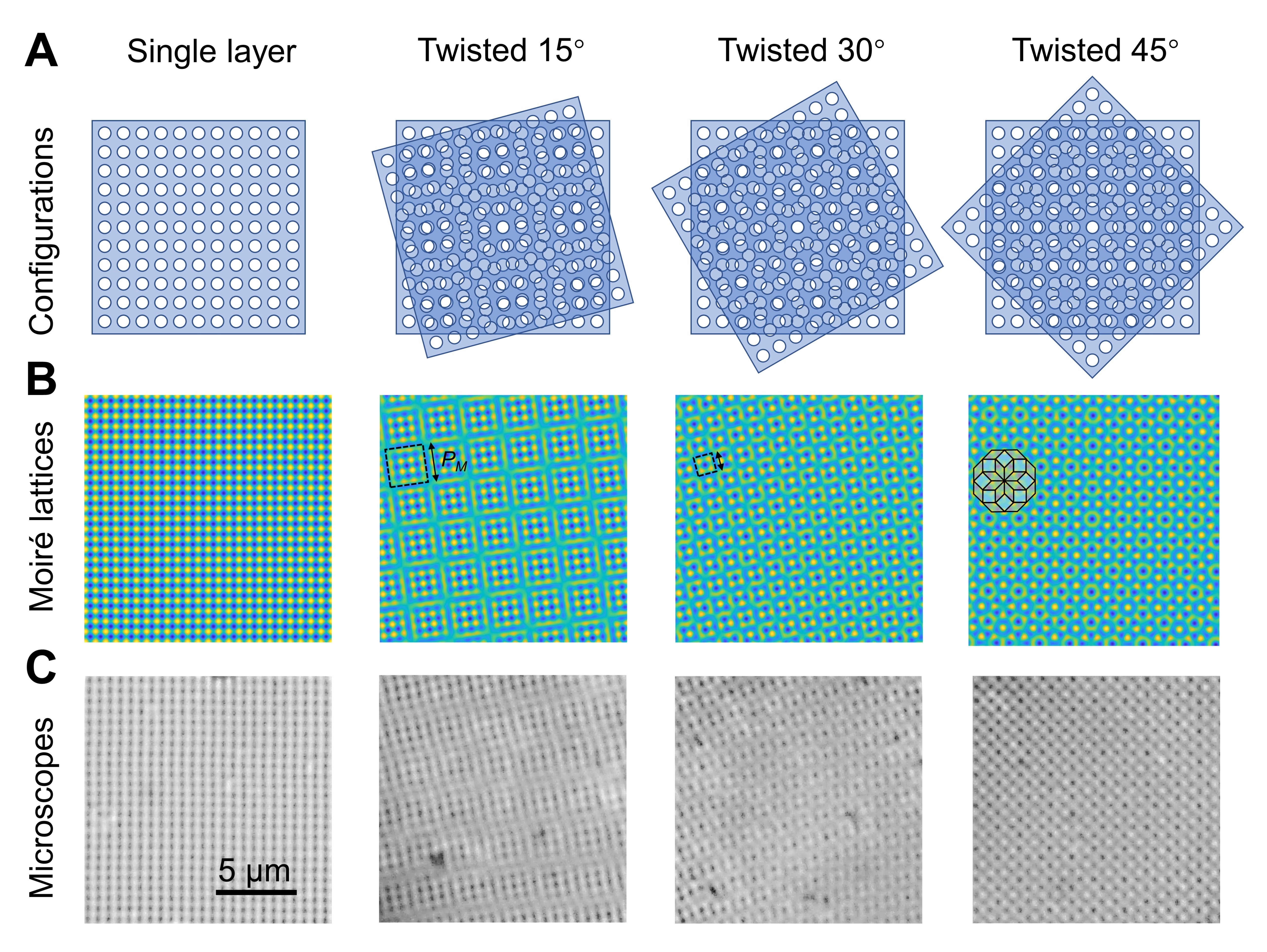}
  \caption{Moir\'e lattices in metasurfaces. a) Schematic representations delineating the formation of moir\'e lattices. The series includes a depiction of a single layer metasurface and three stacked moir\'e metasurfaces with varying twisted angles (15$^\circ$, 30$^\circ$, and 45$^\circ$) arranged from left to right. b) Theoretical calculations of moir\'e lattice patterns. Notably, the 45$^\circ$ moir\'e pattern features 8-fold symmetry. An inset showing the corresponding Ammann-Beenker tiles is presented in the top-left corner, providing additional insight into the lattice structure. c) Experimental observations of moir\'e patterns captured through microscopy, providing a visual confirmation of the constructed metasurfaces.}
 \label{fig3}
\end{figure*}

\textbf{Figure 3} shows the formation of moir\'e lattices through the stacking of metasurfaces. In Figure 3a, we show the configurations, starting with a single layer square lattice in the first panel, and progressing to three moir\'e metasurfaces created by stacking two identical square lattices with twisted angles of 15$^\circ$, 30$^\circ$ and 45$^\circ$, respectively. Figure 3b plots the calculated patterns of both single layer and moir\'e lattices. The stacking process introduces novel quasi-periodical features, notably the enlarged moir\'e squares resulting from 15$^\circ$ and 30$^\circ$ twisting, as highlighted by the dashed black squares. It is imperative to recognize that our design incorporates incommensurate twisted angles, resulting in aperiodic rather than periodical moir\'e patterns. The quasi-period of the moir\'e lattice can be determined using the formula $P_M=P/[2\sin(\alpha/2)]$ (Figure S3 in the Supporting Information).\textsuperscript{\cite{amidror2003moire}} Furthermore, the moir\'e lattice formed with a 45$^\circ$ twisting gives rise to a quasi-crystal lattice, adhering to Ammann-Beenker tiles and exhibiting remarkable 8-fold rotation symmetry.\textsuperscript{\cite{wang1987two, ammann1992aperiodic}} In Figure 3c, optical images of our metasurfaces observed under a microscope are presented. These images closely replicate the theoretical patterns, providing a visual confirmation of the constructed moir\'e metasurfaces.

\subsection{SHG beam shaping by nonlinear moir\'e metasurfaces}

\begin{figure*}[htbp]
  \centering
  \includegraphics[width=160mm]{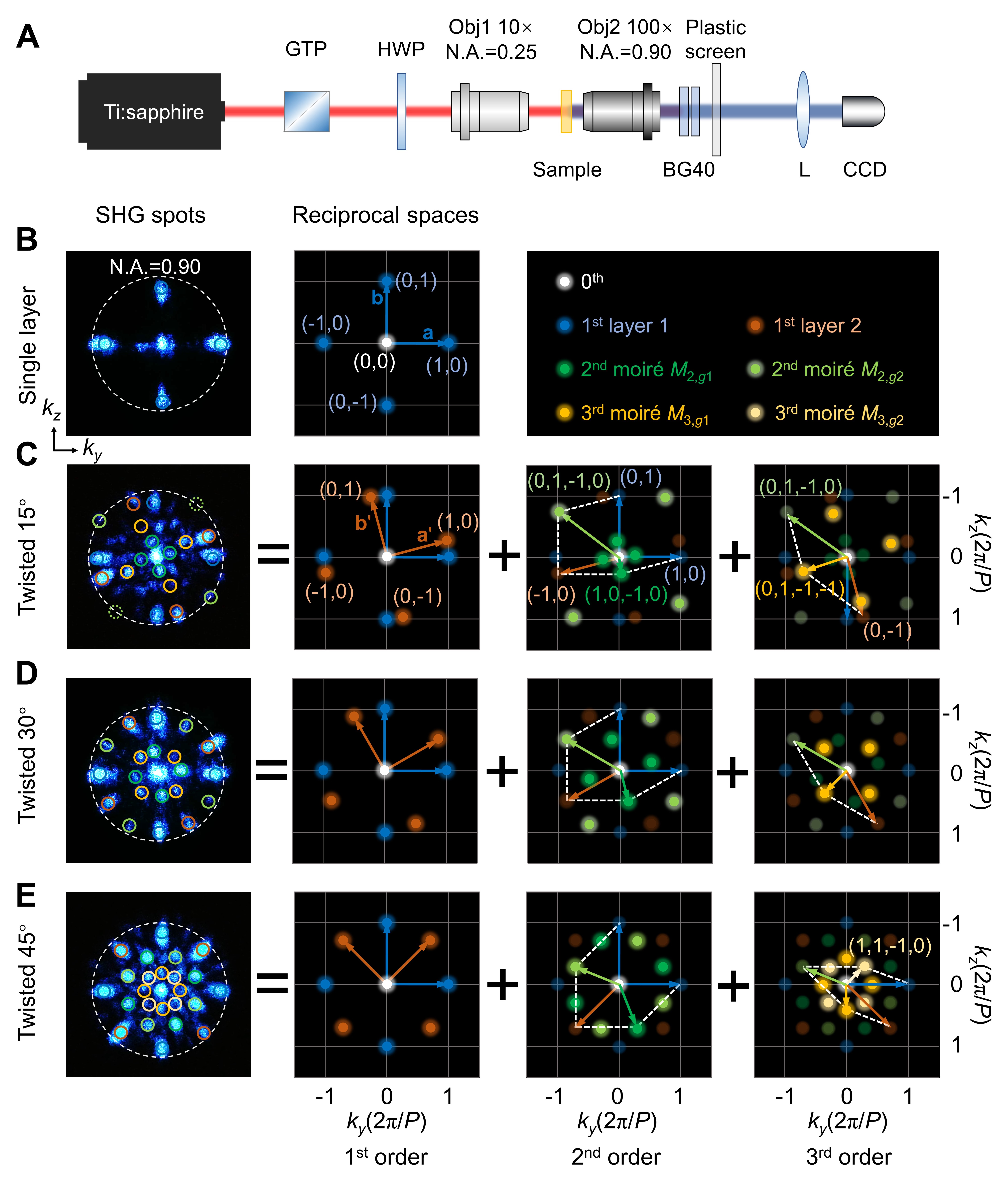}
  \caption{SHG beam shaping by nonlinear moir\'e metasurfaces. a) Experimental optical setup featuring key components such as the Glan-Taylor prism (GTP), half-wave plate (HWP), objective lens (Obj), short-pass filter (BG40), plastic screen, lens (L), and camera. b) The first panel displays the measured SHG pattern from a single layer metasurface, with the dashed circle indicating the N.A. of the objective. The second panel illustrates the reciprocal lattice corresponding to the single layer metasurface, where momentum transfer from the reciprocal lattice to SHG photons generates peripheral first order diffraction spots, marked by blue circles in the first panel. c-e) SHG patterns and reciprocal lattices for different twisted moir\'e metasurfaces. The first column presents the measured SHG patterns, while the second column shows the reciprocal lattices of the untwisted (blue) and twisted (red) metasurface layers, contributing to the SHG spots marked by the blue and red circles in the first column. The third and fourth columns depict reciprocal lattice vectors of the moir\'e metasurface, formed by combining reciprocal lattice vectors of the elementary metasurfaces. These vectors are categorized into several groups, such as $M_{2,g1}$ (dark green), $M_{2,g2}$ (light green), $M_{3,g1}$ (dark yellow), and $M_{3,g2}$ (light yellow), based on different vector combination configurations. This leads to the formation of different groups of SHG spots in the first column, shown by circles of the corresponding colors.}
  \label{fig4}
\end{figure*}

To experimentally characterize the SHG radiation of various moir\'e metasurfaces, the samples were illuminated by a tunable Ti:sapphire femtosecond laser (Maitai, Spectra-Physics, 80~MHz, 230~fs), as shown in \textbf{Figure 4}a. The pump laser was focused onto the metasurfaces by a 10$\times$ objective (N.A.=0.25) forming a focal spot with a diameter of about 10~$\mu$m. A 100$\times$ objective (N.A.=0.90) was used on the transmission side to collect the SHG signals. The transmitted pump wave underwent filtering with short-pass filters (BG40 colored glass). Subsequently, SHG patterns were projected on a semi-transparent plastic screen and captured by a camera. In the experiment, the pump wavelength was selected at 950~nm, overlapping with the transmission resonance of the metasurfaces (Figure S4, Supporting Information). The choice of this wavelength enables resonantly enhanced SHG efficiency.\textsuperscript{\cite{qu2022giant}} Notably, the corresponding SHG wavelength of 475~nm is smaller than the periods of both the elementary metasurface lattice and moir\'e lattice, facilitating diffraction phenomena for the SHG light. Given the short interaction length of the beams in the studied metasurfaces, the diffraction of the SHG is in the Raman-Nath regime. In this regime, lattice momentum transfer along the transverse direction dominates, modulating the propagation direction of the SHG waves.

Figure 4b presents the far-field SHG radiation pattern from the single layer metasurface, with the numerical aperture (N.A.) of the objective delineated by a dashed circle. The normally incident FF light mainly generates SHG photons with zero transverse momentum, resulting in a prominent central spot observed in the pattern. Four additional peripheral spots (marked by four blue circles) are also discernible in the far-field, corresponding to non-zero transverse momentum. The presence of these peripheral SHG spots indicates momentum transfer from the reciprocal lattice to the SHG photons. Specifically, for the square lattice under consideration, the reciprocal lattice manifests as a square array, expressed as $\mathbf{G}_{1,(k,l)}=\frac{2\pi}{P}(k\mathbf{a}+l\mathbf{b})$, where $k$ and $l$ are integers, and $\mathbf{a}$ and $\mathbf{b}$ are unit vectors in reciprocal space. In the second panel of Figure 4b, the reciprocal space of the metasurface is presented. The interaction between the SHG wave and the zero reciprocal lattice vector $\mathbf{G}_{1,(0,0)}$ (i.e., the origin of the reciprocal space) results in the central SHG spot in experiment, as mentioned earlier. Concurrently, the momentum transfer originating from the reciprocal lattice vectors $\mathbf{G}_{1,(\pm1,0)}$ and $\mathbf{G}_{1,(0,\pm1)}$, resulting in the four peripheral spots, i.e., the first order diffraction.

\begin{figure*}[htbp]
  \centering
  \includegraphics[width=180mm]{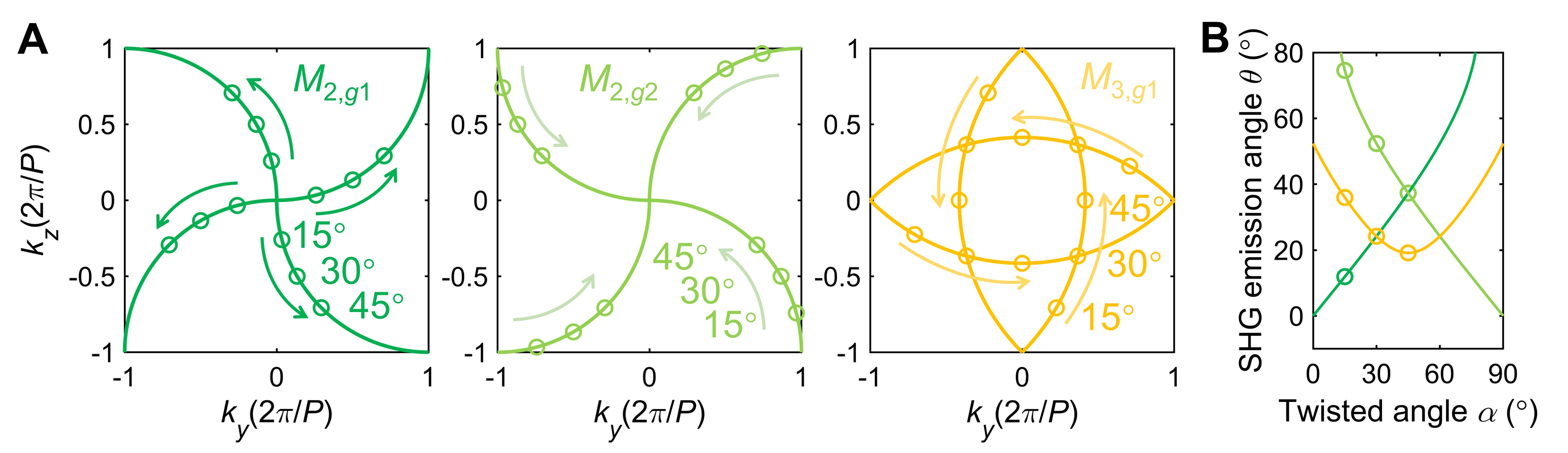}
  \caption{Evolution of SHG spot trajectories and angular correlation with twist angle. a) Trajectories of SHG spots in reciprocal space for groups of $M_{2,g1}$, $M_{2,g2}$, and $M_{3,g1}$, illustrating how their motion changes with the twisted angle $\alpha$. Arrows indicate the direction of increasing $\alpha$ from 0 to 90$^{\circ}$. b) Correlation between SHG propagation angle $\theta$ and the twisted angles $\alpha$ of the moir\'e metasurfaces.}
  \label{fig5}
\end{figure*}

When two metasurfaces are stacked, the interlayer coupling introduces a superposition of reciprocal lattices from the individual metasurfaces. The reciprocal lattice of the twisted layer is expressed as $\mathbf{G}_{2,(m,n)}=\frac{2\pi}{P}(m\mathbf{a}'+n\mathbf{b}')$, where $\mathbf{a}'=\cos(\alpha)\mathbf{a}-\sin(\alpha)\mathbf{b}$ and $\mathbf{b}'=\sin(\alpha)\mathbf{a}+\cos(\alpha)\mathbf{b}$ are unit vectors in the reciprocal space of the twisted metasurface. Consequently, the reciprocal lattice of moir\'e metasurfaces, which is the sum of the reciprocal lattices of each layer, can be represented as $\mathbf{G}_{M,(k,l,m,n)}=\frac{2\pi}{P}(k\mathbf{a}+l\mathbf{b}+m\mathbf{a}'+n\mathbf{b}')$.

Figure 4c illustrates the SHG pattern for a 15$^\circ$ twisted moir\'e metasurface. Beyond the central zeroth order diffraction spot, we identify two distinct groups of first order peripheral spots, distinguished by blue and red circles. These groups exhibit a relative 15$^\circ$ angular offset, each corresponding to the intrinsic emission from either the bottom or top layers. Visual representations of the reciprocal lattices of the elementary metasurfaces are provided in the second panel of Figure 4c, where red and blue dots signify the reciprocal lattices associated with the twisted or untwisted layers, respectively. In contrast to the single layered metasurface, which exclusively features zeroth and first diffraction orders, the twisted moir\'e metasurface exhibits additional SHG spots. The generation of these spots arises from the composition of reciprocal lattice vectors from two elementary lattices. As exemplified in the third panel of Figure 4c, the combination of vectors $\mathbf{G}_{1,(1,0)}$ (horizontal blue arrow) and $\mathbf{G}_{2,(-1,0)}$ (red arrow) results in a vector of $\mathbf{G}_{M,(1,0,-1,0)}$, as indicated by a short dark green arrow. These dark green points in the reciprocal lattice basically originate from the moir\'e lattice, as indicated in Figure 3b. The momentum transfer from these vectors forms the SHG spots closely surrounding the zeroth order spot, indicated by dark green circles in the first panel of Figure 4c. This group of spots is classified as $M_{2,g1}$. Similarly, the combination of vectors $\mathbf{G}_{1,(0,1)}$ (vertical blue arrow) and $\mathbf{G}_{2,(-1,0)}$ generates a vector at $\mathbf{G}_{M,(0,1,-1,0)}$, as indicated by a light green arrow. This contributes to the formation of SHG spots marked by light green circles in the first panel of Figure 4c, labeled as $M_{2,g2}$. Due to the proximity of these spots to the edge of the objective aperture, two spots were missing in the experiment, as indicated by dotted circles in Figure 4c. These spots, resulting from the combination of two reciprocal lattice unit vectors, are designated as second-order diffraction. Additionally, the vector $\mathbf{G}_{M,(0,1,-1,0)}$ combines with $\mathbf{G}_{2,(0,-1)}$ to yield a vector $\mathbf{G}_{M,(0,1,-1,-1)}$ (dark yellow arrows). This forms a group of third-order SHG diffraction spots, which are marked by dark yellow circles in Figure 4c and labeled as $M_{3,g1}$. These sharp diffraction patterns confirm the existence of long-range order in the quasi-periodic moir\'e metasurface. The presence of second and third order diffractions underscores the contribution from the moir\'e lattice momentum to the observed SHG phenomena.

Figure 4d and Figure 4e depict SHG from 30$^\circ$ and 45$^\circ$ twisted moir\'e metasurfaces, including components spanning from zeroth to third order diffractions. The visual representation reveals the pronounced influence of the twist angle on both the position of SHG spots and the symmetry of SHG patterns. Significantly, in contrast to the 15$^\circ$ and 30$^\circ$ twisted moir\'e metasurfaces, where a singular set of third order diffraction spots is observed, the 45$^\circ$ twisted moir\'e metasurface manifests two distinct sets of third order diffraction spots, which are distinguished by dark ($M_{3,g1}$) and light yellow ($M_{3,g2}$) circles in the first panel of Figure 4e. This phenomenon indicates a strong variation in higher-order diffraction characteristics associated with the twist angle. Moreover, owing to the inherent 8-fold quasicrystal nature intrinsic to the 45$^\circ$ twisted moir\'e metasurface (as shown in Figure 3b), the SHG pattern exhibits an octagonal symmetry. This distinctive 8-fold symmetry further elucidates the unique geometric properties introduced by the specific twist angle, adding a layer of complexity to the observed SHG phenomena.

To intuitively elucidate the intricate modulation of SHG emission induced by moir\'e twisting, \textbf{Figure 5} delineates the trajectories of SHG spots as a function of the twist angle. The first panel tracks the evolution of the spot group $M_{2,g1}$ (indicated by dark green in Figure 4). With an increasing twist angle, the SHG spots migrate outward from the origin of the reciprocal space to the corners, following quarter arcs (see Figure S5, Supporting Information). In the second panel, the $M_{2,g2}$ spots traverse quarter arcs as well, but their trajectories, in conjunction with those of $M_{2,g1}$, manifest mirror symmetry about the diagonals of the reciprocal lattice. Intriguingly, these spots shift inward from the corners towards the center with an increasing twist angle, adding a distinctive feature to the SHG emission dynamics. The third panel unveils the trajectories of $M_{3,g1}$ spots. As the twisted angle varies, these spots trace a unique path of quatrefoil, embodying the complex interplay of moir\'e-induced effects on SHG emission. The intricate variation of these trajectories provides a tangible evidence of how moir\'e twisting intricately influences the spatial distribution of SHG spots in a deterministic yet efficient way.

Figure 5b depicts the divergence angles $\theta$ of the SHG spots, which are defined as the angles between the SHG beam path and the normal of the metasurface (as presented in Figure 1). Intersection points between different curves signify SHG spots with identical divergence angles, indicating an equal distance from the central zeroth order. Notably, for the 30$^\circ$ twisted moir\'e metasurface, all the $M_{2,g1}$ and $M_{3,g1}$ spots maintain an equal distance from the canter, as evidenced  in Figure 4d. Additionally, in the case of the 45$^\circ$ twisted moir\'e metasurface, the $M_{2,g1}$ and $M_{2,g2}$ spots form a circular arrangement, as depicted in Figure 4e. These observations provide a concrete demonstration of the opportunities emerging from modulating SHG processes through moir\'e metasurfaces. Furthermore, it is worth noticing that the moir\'e photonics exhibits remarkable versatility and adaptability, allowing for the integration of diverse materials, geometries, and functionalities within the moir\'e lattice framework. Through strategic adjustments in lattice constants, lattice types, and the number of stacked layers, a myriad of complex moir\'e patterns can be generated, each showing a unique evolution. Such versatility endows considerable flexibility in shaping radiation patterns and engineering beam tracking, facilitating a broad spectrum of applications in photonics.

\section{Conclusion}
In summary, our investigation shows the potential of moir\'e metasurfaces to govern nonlinear optical processes, with a specific emphasis on tailored shaping of SHG waves in momentum space. We achieved these features through the fabrication of stacked nonlinear moir\'e metasurfaces employing LN membranes. These metasurfaces, characterized by extended quasi-periodic patterns, give rise to intricate diffracted spots and twist-angle-controlled directional SHG. Our findings highlight the manipulation of angular emission and nonlinear beam patterns through moir\'e twisting. Notably, the efficient modulations of SHG emission patterns through varying twist angles present opportunities to harness the unique properties of moir\'e lattices for applications such as beam steering, beam splitting, light modulation, multi-channel optical communication, and information processing. Such versatile functionalities underscore the broad potential of moir\'e metasurfaces in advancing diverse fields within photonics and nonlinear optics.


\section{Experimental Section}
\subsection{Fabrication of freestanding LN membrane.} The freestanding LN membranes were created by wet etching LN on insulator (LNOI) wafers (NANOLN Co. Ltd.), which consist of a 210~nm thick $x$-cut single crystalline LN film bonded to a 300~$\mu$m thick silicon (Si) substrate with a 2~$\mu$m thick silicon dioxide (SiO$_2$) buffer layer. A 500~nm thick silicon nitride (SiN) film was deposited by inductively coupled plasma chemical vapor deposition (ICPCVD) onto the backside of the Si substrate as a protective layer. UV lithography was used to create patterned resist windows, and reactive ion etching (RIE) was employed to remove the SiN within the windows. Subsequently, the sample was immersed in a potassium hydroxide (KOH) solution, which etched the Si layer within the windows at a rate of approximately 1~$\mu$m/min, halting at the SiO$_2$ layer. The sample was further treated with a buffered oxide etching solution (BOE, NH$_4$F:HF = 20:1) to remove the SiO$_2$ layer within the windows (Figure S1, Supporting Information).\\

\medskip
\textbf{Supporting Information} \par 
Supporting Information is available from the author.

\medskip
\textbf{Acknowledgements} \par 
This work was supported by National Key R\&D Program of China (2023YFA1407200, 2022YFA1404800, 2019YFA0705000); National Natural Science Foundation of China (12222408, 92050114, 12174202, 12074200, 12304423, 12304424); Guangdong Major Project of Basic and Applied Basic Research (2020B0301030009); China Postdoctoral Science Foundation (2022M721719, 2022M711710); 111 Project (B23045); PCSIRT (IRT0149); Fundamental Research Funds for the Central Universities (010-63241512); Simons Foundation. We thank Nanofabrication Platform of Nankai University for fabricating samples.

\medskip
\textbf{Conflict of Interest} \par
The authors declare no conflict of interest.

\medskip
\textbf{Author contributions} \par
L.Q. and W.W. contributed equally to this work. L.Q. and M.R. conceived and performed the design. W.W., L.Q. and L.B. fabricated samples. L.Q. and C.W. performed optical measurements, L.Q., M.R., D.Z., C.L., A.A. and J.X. analyzed data. L.Q., M.R., A.A. and J.X. jointly wrote the manuscript. All authors discussed the results and prepared the manuscript.

\medskip
\textbf{Data Availability Statement} \par
The data that support the findings of this study are available from the corresponding author upon reasonable request.

\medskip

\bibliography{ref}

\end{document}